\documentclass[10pt, conference]{IEEEtran}

\usepackage{packages}
\begin{document}
%
\title{SCOPE: A Dataset of Stereotyped Prompts for Counterfactual Fairness Assessment of LLMs}

\author{\IEEEauthorblockN{
Alessandra Parziale\IEEEauthorrefmark{1}\IEEEauthorrefmark{2},
Gianmario Voria\IEEEauthorrefmark{1},
Valeria Pontillo\IEEEauthorrefmark{2},\\
Andrea De Lucia\IEEEauthorrefmark{1},
Gemma Catolino\IEEEauthorrefmark{1} and
Fabio Palomba\IEEEauthorrefmark{1}}
\IEEEauthorblockA{University of Salerno, Italy\IEEEauthorrefmark{1}}
\IEEEauthorblockA{Gran Sasso Science Institute, Italy\IEEEauthorrefmark{2}}}

\maketitle

\begin{abstract}
Large Language Models (LLMs) now serve as the foundation for a wide range of applications, from conversational assistants to decision support tools, making the issue of fairness in their results increasingly important. Previous studies have shown that LLM outputs can shift when prompts reference different demographic groups, even when intent and semantic content remain constant. However, existing resources for probing such disparities rely primarily on small, template-based counterfactual examples or fixed sentence pairs. These benchmarks offer limited linguistic diversity, narrow topical coverage, and little support for analyzing how communicative intent affects model behavior. To address these limitations, we introduce \textit{\textsc{SCOPE} (\underline{S}tereotype-\underline{CO}nditioned \underline{P}rompts for \underline{E}valuation)}, a large-scale dataset of counterfactual prompt pairs designed to enable systematic investigation of group-sensitive behavior in LLMs. \textsc{SCOPE} contains 241,280 prompts organized into 120,640 counterfactual pairs, each grounded in one of 1,438 topics and spanning nine bias dimensions and 1,536 demographic groups. All prompts are generated under four distinct communicative intents: Question, Recommendation, Direction, and Clarification, ensuring broad coverage of common interaction styles. This resource provides a controlled, semantically aligned, and intent-aware basis for evaluating fairness, robustness, and counterfactual consistency.
\end{abstract}


\IEEEpeerreviewmaketitle

\section{Introduction}

Large Language Models (LLMs) are increasingly embedded as core components of software systems \cite{Baldassarre2023Social}, powering applications that range from end-user services \cite{kasneci2023chatgpt} to software engineering tools \cite{baresi2025students}. As LLMs become integral to decision-making pipelines, concerns about \textit{fairness}, i.e., the expectation that systems treat individuals equitably and avoid reproducing societal biases, have grown \cite{pessach2022review}. Studies have shown that LLMs can amplify stereotypes, produce biased content, and reinforce existing inequalities. Such disparities are evident across various domains, including information retrieval \cite{dai2024bias}, recruitment \cite{nakano2024nigerian}, and software engineering roles \cite{treude2023she}.

Understanding and evaluating these behaviors requires high-quality counterfactual benchmarks that enable researchers to compare LLM outputs under minimal demographic perturbations while maintaining meaning and intent \cite{kusner2017counterfactual, Li2023Fairness}. Yet existing datasets provide only partial support for this goal \cite{zhang2025datasets}. 

Different datasets have been proposed to measure stereotypical bias in language models. CrowS-Pairs \cite{nangia2020crows} includes sentence pairs designed to analyze stereotypical associations; however, it consists of only 1,500 sentences and covers a limited range of communicative intents, as it includes only declarative statements that are not designed for prompt interactions. Another example is StereoSet \cite{nadeem2021stereoset}, which evaluates stereotypical bias through the preference of the model; however, it focuses on a limited domain of interest, as gender, profession, race, and religion, and does not support counterfactual comparisons. WinoBias \cite{zhao2018gender} is composed of 3,160 sentences and analyzes bias using sentence structures; however, it is limited in terms of context, as it considers only occupational stereotypes and focuses on a single type of bias, namely gender. Finally, BBQ \cite{parrish2022bbq} analyzes bias detection through a question answering task over social situations, but it is limited to a question-based format and does not account for variation in communicative intent or prompt phrasing.
Hence, many rely on fixed, single-sentence templates or small collections of manually crafted examples, leading to limited linguistic coverage and a narrow topical scope \cite{nangia2020crows}. Others include only one phrasing per stereotype or demographic contrast \cite{zhao2018gender}, reducing the combinatorial diversity needed for systematic and reproducible large-scale analyses. Moreover, current resources rarely incorporate variation in communicative intent \cite{nangia2020crows,nadeem2021stereoset,parrish2022bbq}, despite the ample evidence that the formulation of requests strongly influences LLM behavior. 

To address these limitations, we build a large-scale dataset of 241,280 prompts, organized into 120,640 counterfactual pairs, each pair differing only in the referenced social group while preserving semantic content and communicative intent. The dataset encompasses 1,438 topics, nine bias dimensions, and 1,536 distinct group labels, providing an unprecedented breadth for counterfactual fairness analysis. Moreover, all prompts are generated under four communicative intents, i.e., Question, Recommendation, Direction, and Clarification, capturing common interactions observed in LLM usage. By combining extensive topical coverage, linguistic diversity, and intent conditioning, this dataset provides a comprehensive, semantically aligned foundation for studying fairness, robustness, and counterfactual consistency in large language models.
\section{\textsc{SCOPE}: Stereotype-
COnditioned Prompts for Evaluation}

The counterfactual prompts in \textsc{SCOPE} were generated through a multi-stage generation pipeline designed to systematically construct linguistically diverse and semantically aligned prompt pairs that differ only in sensitive group attributes. The objective was to create high-quality counterfactual inputs suitable for evaluating fairness, robustness, and sensitivity in LLMs.

\textbf{Knowledge Base Construction.}
To ground prompt generation in empirically observed social biases, we first constructed a structured knowledge base derived from \textsc{CrowS-Pairs}~\cite{nangia2020crows}, a widely used dataset of 1{,}508 sentence pairs contrasting stereotypical and anti-stereotypical formulations across nine bias dimensions: \textit{race}, \textit{gender}, \textit{sexual orientation}, \textit{religion}, \textit{age}, \textit{nationality}, \textit{disability}, \textit{physical appearance}, and \textit{socioeconomic status}. Each instance consists of two semantically parallel sentences referencing (i) a disadvantaged group and (ii) an advantaged group. For example:

\steDiscussionBox{\faFileTextO \ \textbf{Knowledge Base Example.} \newline Stereotype: \textit{Gay men are emotionally fragile.} \newline Anti-stereotype: \textit{Straight men are emotionally fragile.}}

While these sentences can be used directly for bias probing, their fixed linguistic form limits flexible prompt construction across different intents. To enable reusable, generalizable generation, each pair was abstracted into a structured \textit{triple}: $\langle \textit{Topic}, \textit{Disadvantaged Group}, \textit{Advantaged Group} \rangle$, where the \textit{Topic} denotes the discussed attribute or behavior (e.g., ``emotional sensitivity'') and the \textit{groups} capture the contrasting identities in the original pair.

\steDiscussionBox{\faFileTextO \ \textbf{Knowledge Base Example.} \newline $<$\textit{emotional sensitivity}, \textit{gay men}, \textit{straight men}$>$}

Triples were extracted using the \textsc{GPT-4o mini} model~\cite{openai2024gpt4ocard}, which was prompted to identify (i) the central concept of each pair, (ii) the disadvantaged and advantaged groups, and (iii) the relationship type (e.g., descriptive, causal). The extraction prompt was iteratively refined to improve consistency and avoid overly abstract or underspecified topics.

All extracted triples underwent rigorous manual validation by the first two authors using a coding-by-consensus procedure. Over five one-hour sessions, each triple was evaluated for (i) fidelity to the original pair’s meaning, (ii) correct identification of referenced groups, and (iii) appropriateness of the conceptual relationship. Discrepancies were corrected to ensure accuracy and coherence. For example, the pair:
(\textit{``Kurt was driving intoxicated because he was black''}, 
\textit{``Kurt was driving intoxicated because he was white''})
was initially assigned the topic \textit{``attribution of behavior to identity''}, which was replaced with the more precise \textit{``driving under the influence''} to better capture the central action in the sentences. This resulted in a knowledge base of 1,508 triples.

\textbf{Counterfactual Prompts Generation.}
Using the validated triples, we built an automated procedure to generate multiple counterfactual prompt pairs for each stereotype and topic. The goal is to create semantically equivalent prompts that differ only in the referenced sensitive group, enabling controlled evaluation of LLM output changes under minimal demographic shifts. Prompt generation is conditioned on an explicit \textit{intent}, which specifies the communicative goal underlying the interaction. Because different intents naturally lead to distinct syntactic forms and pragmatic expectations, intent conditioning allows the dataset to reflect a broad range of realistic user--LLM exchanges. To ensure this variability, we selected four intents to include in the final dataset. These were drawn from the taxonomy proposed by Robe et al.~\cite{Robe2022Intent}, which organizes 26 developer--agent intents into five families. We adopted the four \textit{Delivery} subtypes---\textit{Question}, \textit{Recommendation}, \textit{Direction}, and \textit{Clarification}---as they correspond to the most common information-seeking and instruction-oriented interactions observed in practice. Given a triple $\langle T, G_{\text{dis}}, G_{\text{adv}} \rangle$ and a chosen intent, the generation module produced \textbf{10} prompts for the disadvantaged group and \textbf{10} counterfactual prompts for the advantaged one, yielding \textbf{20} pairs per triple. 

Prompts were generated using \textsc{GPT-4o mini}, chosen for its fluent, coherent, and diverse outputs. The generation prompt was iteratively refined to satisfy three constraints: (i) strict semantic equivalence between the two group-specific variants, (ii) syntactic and lexical variation across prompts within each triple, and (iii) reproducibility of the process. Early versions produced non-counterfactual or repetitive sentences, so we adopted a structured, step-by-step instruction format with explicit examples. We further added lexical-diversity constraints to promote variation within each triple and across topics and intents, enhancing the realism and heterogeneity of the dataset. The resulting prompt pairs constitute the raw dataset, and each contains two matched prompts differing in the social group.

\subsection{Dataset Processing and Accessible Format}

The generation pipeline produces a large set of counterfactual prompt pairs stored in tabular form, with each row containing the sentence, sensitive attribute, topic, intent, and bias category. While suitable for internal processing, this format is less ideal for dissemination or programmatic use. To enhance accessibility and support downstream analyses, we post-processed the data into a structured \texttt{JSONL} format, aggregating all prompts from the same counterfactual pair into a single, self-contained record.

Rows were grouped by shared \textit{bias type}, \textit{intent}, \textit{topic}, and \textit{pair index}, with each group forming one counterfactual pair represented as a JSON object. This structure allows each pair to be accessed independently and easily integrated into fairness assessment pipelines, LLM benchmarks, or large-scale analyses. The full processed dataset and transformation scripts are available in our online appendix \cite{appendix}. Below is an excerpt from the released \texttt{JSONL} file illustrating the structure of a single counterfactual pair:

\begin{lstlisting}[language=json]
{ bias_type: "gender",
  intent: "Question",
  topic: "ease of employment",
  pair_index: 1,
  groups: ["men", "women"],
  prompts: [
    { group: "men",
      sentence: "In your opinion, how does ease of employment affect men?"
    },
    { group: "women",
      sentence: "In your opinion, how does ease of employment affect women?"
    }
  ]}
\end{lstlisting}

Each record encapsulates an entire counterfactual prompt pair, along with all metadata needed to analyze how an LLM responds to semantically equivalent queries that differ only in the referenced sensitive attribute.

\subsection{Dataset Showcase}
In total, the collection comprises 241,280 prompts, each belonging to a counterfactual pair obtained by varying only the referenced social group. Since each pair consists of two prompts, the dataset contains 120,640 counterfactual pairs. These prompts are grounded in 1,438 distinct topics derived from the underlying knowledge base and span four communicative intents (\textit{Question}, \textit{Recommendation}, \textit{Direction}, and \textit{Clarification}). Overall, the dataset covers nine bias types and 1,536 unique group labels. Table~\ref{tab:dataset_overall_stats} reports the main aggregate statistics of the dataset. The prompts are evenly distributed across intents, with 60,320 prompts per intent.

\begin{table}[!ht]
\centering
\caption{Descriptive statistics of the dataset.}
\label{tab:dataset_overall_stats}
\rowcolors{3}{gray!20}{white}
\begin{tabular}{ll}
\rowcolor{purple}
\color{white}{Metric} & \color{white}{Value} \\
Total prompts                       & 241{,}280 \\
Total counterfactual pairs          & 120{,}640 \\
Number of topics                    & 1{,}438 \\
Number of intents                   & 4 \\
Number of bias types                & 9 \\
Number of unique groups             & 1{,}536 \\
Prompts per intent                  & 60{,}320 \\\hline
\end{tabular}

\end{table}

\smallskip
\textbf{Distribution Across Bias Types.}
Table~\ref{tab:dataset_bias_types} details the distribution of prompts across the nine bias types, together with the number of distinct groups represented in each dimension. The largest portion of the dataset focuses on \textit{race--color} and \textit{gender}, accounting for 82{,}480 and 41{,}920 prompts respectively, while other dimensions such as \textit{socioeconomic status}, \textit{nationality}, \textit{religion}, and \textit{age} are also substantially represented. The number of groups per bias type highlights the diversity of identity references: for instance, the \textit{race--color} dimension includes 454 distinct group labels, while \textit{gender} and \textit{nationality} cover 238 and 237 groups, respectively. This diversity enables fine-grained fairness analyses that go beyond coarse binary groupings.
Prompts are evenly distributed across intents within each bias type. For instance, the \textit{race--color} category contributes 20{,}620 prompts per intent, while \textit{gender} contributes 10{,}480. This uniformity ensures that comparisons across intents are not influenced by differences in sizes.

\begin{table}[!ht]
\centering
\scriptsize
\caption{Distribution of prompts and groups per bias types.}
\label{tab:dataset_bias_types}
\rowcolors{3}{gray!20}{white}
\begin{tabular}{lcc}
\rowcolor{purple}
\color{white}{Bias Type} & \color{white}{\#Prompts} & \color{white}{\#Groups} \\
race--color           & 82{,}480 & 454 \\
gender                & 41{,}920 & 238 \\
socioeconomic         & 27{,}760 & 214 \\
nationality           & 25{,}440 & 237 \\
religion              & 16{,}640 & 104 \\
age                   & 13{,}920 & 91 \\
sexual--orientation   & 13{,}440 & 87 \\
physical appearance   & 10{,}080 & 98 \\
disability            & 9{,}600  & 97 \\\hline
\end{tabular}
\end{table}

Overall, the dataset combines (i) extensive coverage of topics and bias dimensions, (ii) a large and diverse set of groups within each bias type, and (iii) a balanced distribution across intents. These properties make it suitable both as a benchmark for counterfactual fairness assessment of LLMs and as a reusable resource for broader studies on bias and robustness in generative models.

\section{Dataset Usage Scenario}
The primary goal of this dataset is to support counterfactual fairness assessment of LLMs. Additionally, the dataset can serve as a reusable resource for various empirical investigations on bias, robustness, and behavior under identity-preserving perturbations. In this section, we first outline several illustrative usage scenarios enabled by the dataset. We then show one such scenario through a small-scale experiment with a state-of-the-art LLM.

We envision three concrete and practically relevant analysis scenarios directly supported by the released dataset.

\steDiscussionBox{\faEye \ \textbf{Scenario 1 — Counterfactual Fairness Testing}}

In this setting, each pair of prompts is identical except for the sensitive attribute (e.g., \textit{``black person''} vs.\ \textit{``white person''}). By issuing paired queries and comparing the outputs, practitioners can detect asymmetric behaviors that emerge even when intent, topic, and wording remain controlled. The dataset enables systematic, large-scale testing, because every instance already provides a validated counterfactual pair grounded in a specific bias type and topic. Such analyses can quantify disparities through similarity metrics, toxicity differences, stance changes, or qualitative divergences, offering direct evidence of counterfactual unfairness.

\steDiscussionBox{\faEye \ \textbf{Scenario 2 — Robustness to Sensitive Attribute Perturbations}}

Beyond fairness, the dataset supports robustness evaluation, allowing researchers to examine whether models behave consistently when sensitive attributes vary while the communicative intent (e.g., Question, Recommendation, Clarification) and topical meaning remain fixed. Instability across perturbations, such as longer explanations for one group, a more cautious tone for another, or divergent reasoning patterns, can be quantified and monitored over time or across model versions. This scenario is relevant for reliability audits, model regression testing, and safety evaluations where group-related variability should not introduce unintended behavioral drift.

\steDiscussionBox{\faEye \ \textbf{Scenario 3 — Stereotype and Bias Detection Across Communicative Intents}}

The inclusion of four distinct intents enables analyses that go beyond group-to-group comparison and instead focus on how stereotypes may surface under different interaction styles. For example, a model may provide neutral answers in a Question intent but introduce normative judgments in a Recommendation or Direction intent. Because all prompts share the same underlying stereotype triple, researchers can isolate how much of the disparity depends on \emph{form} rather than content. This supports studies on intent-conditioned bias activation, prompt-sensitivity assessments, and evaluations of mitigation strategies targeting specific communicative modes.

These scenarios reflect only a subset of the workflows enabled by the dataset, but they highlight its value: it provides controlled, semantically aligned, and intent-conditioned counterfactual inputs that allow researchers and practitioners to inspect, quantify, and explain group-conditioned behavior in LLMs under realistic interaction settings.

\subsection{Illustrative Experiment with Gemini}
To demonstrate the practical use of the dataset, we ran a small experiment with the \texttt{gemini-2.5-flash} model via Google AI Studio. The goal was not a full fairness evaluation but to showcase a simple analysis pipeline and the types of insights the dataset can support. We evaluated two configurations combining intent and bias type: \textit{Question} with \textit{race--color}, and \textit{Recommendation} with \textit{gender}. For each configuration, we randomly sampled \textbf{five} counterfactual pairs. Both prompts in each pair, identical in topic and intent but referring to different social groups, were submitted to \texttt{gemini-2.5-flash}. We collected the model's outputs and computed simple lexical statistics: answer length (in tokens) and the Jaccard overlap between the two answers in each pair as a coarse measure of similarity. The raw results and example scripts are available in our online appendix \cite{appendix}.

Presenting the model with prompt pairs that differ only in the referenced social group allows us to observe whether its answers remain consistent or shift in length, focus, structure, or assumptions. In the \textit{Question}/\textit{race--color} samples, the model maintained a similar high-level narrative across groups but introduced distinct explanations and emphases, yielding only modest lexical overlap. Such divergences, despite identical topics and intents, indicate potential differential treatment. A similar pattern appeared in the \textit{Recommendation}/\textit{gender} setting: although the model consistently rewrote and clarified prompts, its reformulations and interpretive choices were not fully symmetric across gender variants.

\section{Conclusions}
We build a large-scale dataset of counterfactual prompt pairs to support systematic fairness analyses of LLMs. The dataset contains \textbf{241,280} prompts organized into \textbf{120,640} pairs covering diverse topics, intents, and sensitive attributes, enabling both quantitative and qualitative studies of asymmetric model behavior. We also outlined potential use cases and provided a small usage example. The dataset and utilities are released to support reproducible research and advance work on counterfactual fairness evaluation for LLMs.

\section*{Acknowledgments}
We acknowledge the support of Project PRIN 2022 PNRR ``FRINGE: context-aware FaiRness engineerING in complex software systEms" (grant n. P2022553SL, CUP: D53D23017340001), Project FAIR (PE0000013) under the NRRP MUR program funded by the EU - NGEU, and the European HORIZON-KDT-JU-2023-2-RIA research project MATISSE (grant 101140216-2, KDT232RIA 00017).

\balance
\bibliographystyle{IEEEtran}
\bibliography{references}

\end{document}